\documentclass{emulateapj}
\pdfoutput=1

\usepackage{natbib}

\def\wh                {W~Hz$^{-1}$}

\shorttitle{}
\shortauthors{Giodini et al.}

\begin{document}


\title{Radio galaxy feedback in X--ray selected groups from COSMOS: the effect on the ICM }

\author{S.~Giodini\altaffilmark{1},
V.~Smol\v{c}i\'{c}\altaffilmark{2},
A.~Finoguenov\altaffilmark{1,3},
H.~Boehringer\altaffilmark{1},
L.~B$\mathrm{\hat{i}}$rzan\altaffilmark{5},
G.~Zamorani\altaffilmark{9},
A.~Oklop\v{c}i\'{c}\altaffilmark{6},
D.~Pierini\altaffilmark{1},
G.W.~Pratt\altaffilmark{4},
E.~Schinnerer\altaffilmark{8},
R. Massey \altaffilmark{2},
A.M.~Koekemoer\altaffilmark{7},
M. Salvato \altaffilmark{12},
D.B.~Sanders\altaffilmark{10}
J. S. Kartaltepe\altaffilmark{10}
D. Thompson\altaffilmark{11}}
\altaffiltext{1}{Max Planck Institut f{\"u}r Extraterrestrische Physik, Giessenbachstrasse, Garching bei M{\"u}nchen D-85748, Germany}
\altaffiltext{2}{California Institute of Technology, MC 105-24, 1200 East California Boulevard, Pasadena, CA 91125}
\altaffiltext{3}{University of Maryland, Baltimore County, 1000 Hilltop Circle, Baltimore, MD 21250.}
\altaffiltext{4}{Laboratoire AIM, IRFU/Service dÕAstrophysique - CEA/DSM - CNRS - UniversitŽ Paris Diderot, B‰t. 709, CEA-Saclay, F-91191, Gif-sur-Yvette Cedex, France}
\altaffiltext{5}{Leiden Observatory, Leiden University, PO Box 9513, 2300 RA, Leiden, the Netherlands}
\altaffiltext{6}{University of Zagreb, Physics Department, Bijeni\v{c}ka cesta 32, 10000 Zagreb, Croatia}
\altaffiltext{7}{Space Telescope Science Institute, 3700 San Martin Drive, Baltimore, MD 21218}
\altaffiltext{8}{Max-Planck-Institut fŸr Astronomie, K\"onigstuhl 17, D-69117 Heidelberg, Germany}
 \altaffiltext{9}{INAF Osservatorio Astronomico di Bologna, Via Ranzani 1, I-40127 Bologna, Italy}
\altaffiltext{10}{Institute for Astronomy, University of Hawaii 2680 Woodlawn Drive, Honolulu, HI  96822}
\altaffiltext{11}{Large Binocular Telescope Observatory, University of Arizona, 
Tucson, AZ85721, USA }
\altaffiltext{12}{Excellence Cluster Universe and IPP-Max Planck institute for Plasma Physics
Boltzmann Strasse 2, Garching 85748 Germany}
\begin{abstract}
We quantify the importance of the mechanical energy
released by radio-galaxies inside galaxy groups. 
We use scaling relations to estimate the mechanical energy released by 16 radio-AGN
located inside X-ray detected galaxy groups in the COSMOS field. 
By comparing this energy output to the host groups' gravitational
binding energy, we find that radio galaxies produce sufficient energy to unbind a significant fraction of the intra-group medium. 
This unbinding effect is negligible in massive galaxy clusters with deeper potential wells.
Our results correctly reproduce the breaking of self-similarity observed in the scaling relation between entropy and temperature for galaxy groups.
\end{abstract}


\section{Introduction}
Galaxy groups are important laboratories in which to investigate the importance of
non--gravitational processes in structure formation. 
These processes are potentially more important in galaxy groups than
in massive clusters because of their lower gravitational binding
energy. This is suggested by the significant deviation of the observed
X--ray luminosity and entropy versus temperature ($L_{X}$--T and $S$--T) scaling relations in groups compared to the relation
expected in a purely gravitational scenario
(see also \citealt{pratt03}; \citealt{markevitch98}; \citealt{arnaud99}; \citealt{ponman03}; \citealt{sun09}; \citealt{pratt10}). 
Radiative cooling can be invoked to explain this deviation, but then the 
predicted fraction of stars in clusters of a given mass is incorrect 
\citep{voit05,balogh08}.
To simultaneously explain the properties of
the intra--cluster/group medium (ICM) and account for the observed properties of
galaxies, it is necessary to take into account a major contribution to the
cluster/group energetics from non--gravitational heating. 

The two main sources of non--gravitational heating are star--formation and active galactic nuclei (AGN). Cosmological simulations (e.g. \citealt{kay04}; \citealt{bower06}; \citealt{sijacki06}) show that both processes are required
to reproduce the properties of the ICM. In particular, recent
simulations by \citet{bower08} successfully reproduce both the galaxy
and ICM properties (see \citealt{short08}) when they include a
{ "radio--mode" AGN feedback phase: in this phase the movement of bubbles inflated by the AGN jets transfers energy into the gas within the cluster (mechanical heating)}. The observable
objects providing this type of feedback
inside groups and clusters would be radio galaxies \citep{croton06}.  The main difference
between the \citet{bower08} model and others, including radio--mode AGN,
(\citealt{bower06}; \citealt{sijacki06}; \citealt{puchwein08};) is
that it allows the radio mode feedback to expel gas
from the X-ray emitting regions of the { system}. 

The importance of such AGN-feedback in groups could explain the observational result by   \citet{lin03}, \citet{mccarthy07} and \citeauthor{giodini09} (2009), that the total baryon fraction in groups is lower than the
cosmic value estimated from cosmic microwave background (CMB) observations (see \citealt{giodini09} for more details). The discrepancy decreases in systems of higher total mass, such that it is $<$1$\sigma$ for massive clusters.

In this Paper we propose a simple, direct method to test the hypothesis
that radio galaxies in groups can indeed inject enough mechanical energy to
unbind the intra--cluster gas. 
The Paper is structured as follows. In Section 2 we select a sample of 16 groups from the COSMOS 2 deg$^{2}$ survey discussed in \citet{giodini09}, each hosting a radio galaxy within the
virial radius (\citealt{schinnerer07}; \citealt{smolcic08a}), plus a control sample of massive clusters from \citet{birzan04}. 
In Sections \ref{analysis} and \ref{results} we then compare the groups' binding energy to the mechanical energy output by the radio sources, derived from their total radio luminosity through scaling relations. 
Applying this method,
we show that the mechanical removal of gas from the group region is
indeed energetically feasible for systems below
$\sim$3$\times$10$^{14}$M$_{\odot}$. In Section \ref{discussion} we discuss how this scenario compares to the deviation in the scaling relation between entropy and temperature at the groups scale.

{ We adopt a $\Lambda$CDM cosmology with $h$=0.72, $\Omega _{m}$=0.25, $\Omega_{\Lambda}$=0.75.}

\section{The samples}\label{samples}

\subsection{Radio galaxies in X--ray detected groups}
We use the catalog of 91 X--ray selected groups from the COSMOS survey
(\citealt{scoville07a}; Finoguenov et al.\ in preparation), selected as described in
\citet{giodini09}. Extended source detection was performed using
a multiscale wavelet reconstruction of a mosaic of XMM and Chandra
data.  For each group, member galaxies are identified within R$_{500}$\footnote{{ R$_\mathrm{\Delta}$ ($\Delta$=500,200) is the radius within which the mass density
of a group/cluster is equal to $\Delta$ times the critical density ($\rho_{c}$)
of the Universe. Correspondingly, M$_\mathrm{\Delta}=\Delta\,\rho_{c}(z)\,(4\,\pi/3)R_{\Delta}^3$ is the mass inside R$_\mathrm{\Delta}$.  M$_\mathrm{200}$  is computed using an L$_\mathrm{X}$--M$_\mathrm{200}$ relation established  via the weak lensing analysis in \citealt{leauthaud09}. The catalogue value of M$_\mathrm{200}$ is converted into M$_\mathrm{500}$
assuming an NFW profile with a concentration parameter computed from the mass-dependent relation of \citet{maccio07}.}}
of the group center, utilizing the high quality photometric redshifts
available ($\sigma(\Delta z)/(1+z)$=0.02 at  i$_{AB}\footnote{ AB magnitude in the SUBARU i band.}<$25, \citealt{ilbert09}). 

We use a sub-sample of the VLA--COSMOS catalog
(\citealt{schinnerer07}; \citealt{smolcic08a})  to identify radio galaxies lying inside the
X--ray selected groups. Of the $60$ radio galaxies\footnote{The term ``radio galaxy'' is used
  here to describe an extended radio source with clear jet/lobe
  structure.} identified within the VLA-COSMOS Large
Project (\citealt{schinnerer07}; 1.49~GHz), about 80\% have
been associated with a secure optical counterpart \citep{smolcic08a} with $i_\mathrm{AB}\leq26$, and accurate photometry (thus
also with accurate photometric redshifts; \citealt{ilbert09}; \citealt{salvato09}).

We have cross-correlated this sample of radio galaxies
with the X-ray selected galaxy groups  in 3D
space using a search radius of $1\times R_{200}$ (Finoguenov et al.\ in preparation) around the groups' centers  and within 0.02$\times$(1+z) from the group's redshift. This resulted in a sample of 16 systems matched in position and redshift.
{ In Appendix A we show the contours of the radio 20 cm and X--ray emission superimposed to the SUBARU $zp$ band  image for each of the groups.}
 In 9 out of 16 cases the
radio galaxy is located in the core of the group (defined as
$R<$0.15$R_{200}$). The 20~cm radio luminosity { densities} \footnote{Computed
  using the total flux densities ($F_\nu$). { K--correction is also applied assuming a spectral
  index of $\alpha=0.7$ ($F_\nu\propto\nu^{-\alpha}$).} } of these
galaxies range from $\sim5.5\times10^{22}-4.8\times10^{25}$~\wh , with
a median luminosity of $8.9\times10^{24}$~\wh ($\nu\,F_{\nu}\sim7.3\times10^{38}-6.4\times10^{41}$~erg s$^{-1}$, with
a median luminosity of $1.18\times10^{41}$~erg s$^{-1}$) . This median luminosity is
at the high end of the radio luminosity distribution of the full radio
AGN sample (c.f.\ Fig.~17 in \citealt{smolcic08a}, and Fig.~5
in \citealt{smolcic09}), consistent with previous findings
that powerful radio galaxies inhabit group-scale environments
(e.g. \citealt{baum92}).
The redshift distribution of the 16 groups is
fairly uniform between 0.1 and 1, with the exception of 6 sources
concentrated at z$\sim$0.3 (where a large structure extends throughout the
whole COSMOS field). The groups have X--ray luminosities ranging from
1$\times$10$^{42}$ to 8.7$\times$10$^{43}$~erg~s$^{-1}$ and span a
mass range of 2$\times$10$^{13}<$M$_{200}<$2$\times$10$^{14}$M$_{\odot}$ with a median mass of  7.14$\times$10$^{13}$M$_{\odot}$.

\subsection{The comparison sample of massive clusters}\label{birzan}
The COSMOS X--ray sample is mostly composed of groups. We complement it with 
12 well known radio galaxies inside massive clusters, extracted from the sample of \citet{birzan04}. 
{  We use those clusters from the Birzan's sample which overlap with the HIFLUGCS survey \citep{reiprich02} so that we can use the X--ray parameters determined from the HIFLUGCS clusters. In addition we require that the radio source within those clusters is associated with a secure NIR counterpart in the 2MASS catalogue \citep{2mass}. These requirements eliminate 4 of the clusters in the original Birzan et al. sample.}
Each of these clusters contain X--ray cavities associated with radio bubbles likely connected with AGN activity of the central galaxy. The radio galaxies have been identified within the NRAO VLA Sky Survey (NVSS) at 1.49 GHz \citep{condon98}, except in the case of the Centaurus Cluster where data come from the 1.41 GHz Parkes Radio Sources Catalogue \citep{wright90}. The 20 cm radio luminosities of the radio galaxies range between  2$\times$10$^{39}$ and 2 $\times 10^{43}$ erg s$^{-1}$, with a median luminosity of 1.4$\times 10^{42}$ erg s$^{-1}$,  { more than 10 times higher than the median radio luminosity of the radio galaxies in the COSMOS sample}.\\
The X--ray parameters for these clusters are provided by the X--ray analysis in the HIFLUGCS survey, and converted for the standard cosmology used in this paper.
The sample consists of very local clusters, ranging in total mass between 1$\times$10$^{14}<$M$_{200}<$1.2$\times$10$^{15}$M$_{\odot}$  with a median mass of 4.25$\times$10$^{14}$M$_{\odot}$,  { almost 10 times higher than the median total mass of the systems in the COSMOS sample}\\

\section{Analysis of COSMOS group sample}\label{analysis}
\subsection{Mechanical energy input by radio galaxies in groups}
We estimate the mechanical energy input by a radio galaxy into the ICM { over the group lifetime} from the mechanical luminosity of the radio source multiplied by the fraction of time a
massive galaxy spends in the radio-AGN phase.  The mechanical
luminosity for the radio galaxies in our sample is estimated from the
scaling relation presented in \citet{birzan08}.  These authors studied
a sample of galaxy clusters showing signatures of cavities and bubbles
in the X-ray surface brightness 2D distribution, with a powerful radio source
as a central galaxy. 
The cavity power of the radio source, estimated from the $p{\mathrm d}V$ work of the jet/lobe on the surrounding ICM, is found to be correlated (albeit with a large scatter) with the
monochromatic radio power at 1.49 GHz of the central galaxies ($P_{1.49\,GHz}$) as
\begin{equation}
 P_{cav}\propto P_{1.49\,GHz}^{0.35 \pm 0.07}
\end{equation}
(see Eq. 16 in \citealt{birzan08}). $P_{1.49\,GHz}$ is computed from the radio--emission of the 
{ entire source}. 
This estimate is a lower limit to the mechanical luminosity of the
{ AGN outbursts}, since it does not take into account the energy dissipated 
(e.g. in shocks). $P_\mathrm{cav}$ is related to the $p{\mathrm d}V$ work through 
\begin{equation}
 P_{cav}= \frac{4PV}{\tau}
\end{equation}
(\citealt{churazov02}; \citealt{birzan08}), where $\tau$ is the { duration of each single AGN outburst and 4 is the factor used for relativistic plasma}.
\citet{smolcic09} investigated the fraction of
radio AGN as a function of cosmic time and stellar mass of
the galaxy. This fraction can be related, through a probability
argument detailed in \citet{smolcic09}, with the time a galaxy of a given stellar mass and at a given redshift spends as a radio galaxy ($\tau_\mathrm{radio}$).  Using this result
we can estimate the average duration of radio sources as a
function of redshift and stellar mass of the host galaxy (see Fig.~12 in \citealt{smolcic09})\footnote{To derive the values of $\tau_\mathrm{radio}$ \citet{smolcic09}, it is assumed that the
radio parent population (red massive galaxies) is formed at $z=3$ \citep{renzini06} and
survives until $z=0$. Since the COSMOS radio galaxies are not at $z=0$, the time-scales computed in  \citet{smolcic09} coincide with ours if multiplied by $\frac{t(z=zgal)-t(z=3.0)}{10^9\,yr}$, where $t$ is the age of the universe at redshift $z$ and $zgal$ is the redshift of the radio galaxy.}. This gives a plausible time-scale during which the radio
AGN can have injected mechanical energy into its environment. 
 For the 16 COSMOS X--ray selected groups, $\tau_{radio}$  ranges between  0.003 and 4.18 Gyr, with a median value of 3.1 Gyr.
The mechanical energy contribution
can then be estimated as
\begin{equation}\label{emech}
E_\mathrm{mech}=P_\mathrm{cav}\times \tau_\mathrm{radio}.
\end{equation}
The values of E$_\mathrm{mech}$ for our sources are shown in Table 1 and span a range between $\sim$2$\times$10$^{57}$--3$\times$10$^{61}$ erg h$_{72}^{-2}$.
The uncertainties in the radio mechanical energy input are dominated by the scatter in the
scaling relation used to convert the monochromatic power into
mechanical luminosity, which amounts to 0.85 dex, and by the uncertainties on $\tau_\mathrm{radio}$. We use $\tau_\mathrm{radio}$ as derived from 
an average estimate over a sample of radio galaxies in the COSMOS field as
a whole, irrespective of their environment. One might
expect the density of the environment surrounding the jets to have a
significant impact on the jet lifetime. However, the fraction of radio galaxies that resides within  the COSMOS groups is comparable with the fraction of red massive galaxies within groups in the control sample used in  \citet{smolcic09} (respectively 18$\%$  and 16$\%$ within $R_{200}$); this assures that the statistical argument used to compute the time--scales holds also in this case. Furthermore we can estimate an average time--scale based on only extended radio galaxies in the whole COSMOS group sample as follows. Of the 141 COSMOS groups at $z<1$ and with L$_{X}>10^{42}$ erg s$^{-1}$, 32 contain  a multi-component radio galaxy. Therefore the average duration of the radio galaxy activity during this time interval is $(32/141)\times (t(z=1)-t(z=0))$. This is $\sim1.7$~Gyr, a time-scale comparable with the average life-time estimated with the method by \citet{smolcic09} ($\sim$1.6 Gyr\footnote{computed as $\frac{max(\tau_{radio})-min(\tau_{radio})}{2}$)}.

\begin{center}
 \begin{deluxetable*}{rccccccccc}
 \tabletypesize{\scriptsize}
\tablecaption{{ For each of the 16 COSMOS groups}, the columns indicate 1) X--ray catalogue ID number 2) R.A. 3) Dec. 4) redshift 5) Power at 1.4 GHz 6) Mechanical Power  7)Binding energy $\pm$ 1$\sigma$ confidence limit 8) $E_{mech}$ $\pm$ 1$\sigma$ confidence limit   9) $\tau_\mathrm{radio}$ 10) distance from the center
 \label{tab1}}
\tablewidth{0pt}
 \tablehead{
 \colhead{XID}&
 \colhead{R.A.}&
 \colhead{DEC.}&
 \colhead{$z$ }&
 \colhead{P$_{1.49\,GHz}$}&
 \colhead{P$_{cav}$}&
 \colhead{$E_\mathrm{binding}$}&
 \colhead{$E_\mathrm{mech}$ }&
 \colhead{ $\tau_\mathrm{radio}$ }&
 \colhead{$R/R_{200}$}\\
\colhead{} &
\colhead{[J2000]}&
\colhead{[J2000]}&
\colhead{}&
\colhead{[$10^{24}$ W/Hz]}&
\colhead{[$10^{36}$ W]}&
\colhead{ [10$^{60}$erg] }&
\colhead{[$10^{60}$ erg] }&
\colhead{[Gyr] } &
\colhead{}} 
\startdata
107 &        149.60965 & 2.14799 & 0.28 &  1.11 & 7.353 & 5.916$_{1.937}^{2.342}$ & 0.514$_{0.441}^{3.087}$ &  0.221 & 0.3783 \\ [0.2cm]
262 &        149.60007 & 2.82118 & 0.34 &  19.0 & 19.85 & 127.9$_{26.49}^{38.80}$ & 24.15$_{20.70}^{144.9}$ &  3.858 & 0.0011 \\ [0.2cm]
253 &        149.75626 & 2.79472 & 0.49 &  6.71 & 13.78 & 40.16$_{9.721}^{12.26}$ & 10.14$_{8.693}^{60.85}$ &  2.332 & 0.0042 \\ [0.2cm]
246 &        149.76132 & 2.92909 & 0.34 &  0.90 & 6.828 & 289.8$_{67.46}^{76.73}$ & 7.003$_{6.003}^{42.02}$ &  3.252 & 0.6181 \\ [0.2cm]
311 &        149.93796 & 2.60627 & 0.34 &  6.38 & 13.54 & 17.27$_{3.907}^{4.792}$ & 0.183$_{0.157}^{1.101}$ &  0.042 & 0.2195 \\ [0.2cm]
264 &        149.99847 & 2.76914 & 0.16 &  0.32 & 4.775 & 6.940$_{2.247}^{3.179}$ & 0.004$_{0.003}^{0.027}$ &  0.003 & 0.0007 \\ [0.2cm]
281 &        150.08617 & 2.53141 & 0.88 &  8.90 & 15.21 & 84.94$_{22.45}^{27.05}$ & 1.241$_{1.064}^{7.450}$ &  0.258 & 0.8614 \\ [0.2cm]
191 &        150.11434 & 2.35651 & 0.22 &  1.71 & 8.554 & 8.380$_{2.150}^{2.959}$ & 6.122$_{5.248}^{36.73}$ &  2.269 & 0.0757 \\ [0.2cm]
237 &        150.11774 & 2.68425 & 0.34 &  27.7 & 22.65 & 105.1$_{23.49}^{31.65}$ & 19.56$_{16.77}^{117.4}$ &  2.738 & 0.0027 \\ [0.2cm]
29 &        150.17996 & 1.76887 & 0.34 &  30.0 & 23.29 & 58.49$_{12.77}^{16.80}$ & 31.63$_{27.11}^{189.8}$ &  4.306 & 0.0016 \\ [0.2cm]
64 &        150.19829 & 1.98628 & 0.43 &  12.4 & 17.11 & 21.10$_{4.941}^{7.358}$ & 18.06$_{15.48}^{108.4}$ &  3.348 & 0.0030 \\ [0.2cm]
35 &        150.20661 & 1.82327 & 0.52 &  10.2 & 16.00 & 30.87$_{8.342}^{11.07}$ & 15.32$_{13.13}^{91.95}$ &  3.037 & 0.0008 \\ [0.2cm]
6 &        150.28821 & 1.55571 & 0.36 &  1.13 & 7.401 & 77.91$_{17.35}^{20.23}$ & 0.070$_{0.060}^{0.425}$ &  0.030 & 0.2279 \\ [0.2cm]
149 &        150.41566 & 2.43020 & 0.12 &  0.05 & 2.564 & 50.31$_{10.32}^{14.98}$ & 0.002$_{0.002}^{0.014}$ &  0.003 & 0.1957 \\ [0.2cm]
40 &        150.41386 & 1.84759 & 0.96 &  48.5 & 27.54 & 108.1$_{24.40}^{34.85}$ & 1.786$_{1.530}^{10.71}$ &  0.205 & 0.4888 \\ [0.2cm]
120 &        150.50502 & 2.22506 & 0.83 &  16.4 & 18.88 & 425.0$_{90.06}^{112.5}$ & 24.92$_{21.36}^{149.5}$ &  4.185 & 0.0267 \\ [0.2cm]
 \enddata
 \end{deluxetable*}
\end{center}

%



\subsection{Binding energy of the intra--group medium}\label{binding}
We consider the shape of the dark matter halos to be characterized by NFW \citep{navarro96} radial profiles
\begin{equation}
\rho(x)=\frac{\rho_\mathrm{crit}\,\delta_c}{x\,(x+1)^2}
\end{equation}
where $x=r/r_{s}$, $r_{s}$ is the characteristic radius, and $\rho_{c}$ is the critical density of closure of the universe. $\delta_{c}$ is defined as
\begin{equation}
\delta_{c}=\frac{200\,c^3}{3\,\ln(1+c)-c/(1+c)}
\end{equation}
and $c$ is the concentration of the halo. The scale radius and the concentration are linked by the relation $r_{s}=R_{500}/c_{500}$, where $c_{500}$ is the dark matter concentration inside $R_{500}$.
We estimate the binding energy out to $R_{500}$ because the kinetic energy of the infall velocity field  along filaments becomes important beyond this radius \citep{evrard96} and our simple model may not then be applicable. Furthermore, we can evaluate reliable gas masses from the X--ray observations only within $R_{500}$. 
For simplicity, we assume that the gas follows the same distribution as the dark matter.
We define as binding energy the total potential energy needed to push the ICM gas inside $R_{500}$ beyond $R_{200}$. The binding energy is computed as 
\begin{eqnarray}\label{ebin}
E_\mathrm{binding}&=&\int^{M_{g,500}}_{0}\left[\phi(r)-\phi(R_{200})\right]\,dM_{g} \nonumber \\
&=&4\,\pi \int^{R_{500}}_{0}\phi(r)\,\rho_{g}(r)\,r^2\,dr  
\end{eqnarray}

We neglect the additive constant given by the term $\phi(R_{200})$, as it is small with respect to the other terms of the equation. We use the definition of gas mass within R$_{500}$ as  
\begin{equation}\label{mgas}
M_{g}=M_{g}(R_{500})=4\,\pi\,\int^{R_{500}}_{0}\rho_{g}(r)\,r^{2}\,dr.
\end{equation}
The potential of a spherical NFW model is \citep{hayashi07}
\begin{equation}
\phi(r)=A\times\frac{ln(1+x)}{x}
\end{equation}
where A is
\begin{equation}
A=-\frac{G\,M_{200}}{r_{s}\,(ln(1+c)-c/(1+c))}
\end{equation}
Thus, substituting the terms into Equation~\ref{ebin}, we compute the binding energy of the ICM gas in a NFW dark matter halo as follows:
\begin{equation}
E_\mathrm{binding}=  f_{gas}4\,\pi\,\rho_{crit}\,\delta_c\,A\,r_{s}^3\int^{c_{500}}_{0}\frac{ln(1+x)}{(1+x)^2}\,dx
\end{equation}
where $f_\mathrm{gas}$ is the gas fraction. 
The concentration parameter for the COSMOS groups has been computed from the mass-dependent relation of \citet{maccio07}.
The errors bars on the binding energy are estimated using a Monte Carlo method to numerically propagate the errors on $M_{200}$ and $R_{200}$, the scatter in the $c$--$M_{200}$  and in the $f_\mathrm{gas}$--$M_{500}$ relation.\\
We cannot estimate the gas masses from most of the existing X-ray  observations of the COSMOS X--ray selected groups because of insufficient signal-to-noise. We therefore estimate the gas fraction in the groups from the mean trend of the gas mass
fraction as a function of M$_{500}$. 
This trend was established from an
{ independent compilation of high quality observations of local (z$<$0.2) groups and clusters in the same mass range as the sample under consideration here \citep{pratt09}.}
The observed relation ($f_\mathrm{gas}\propto$M$_{500}^{0.21}$) suggests that lower mass systems
have proportionally less gas than high mass systems.

\section{Analysis of galaxy cluster sample}
\subsection{Mechanical energy input by radio galaxies in massive clusters}

In order to compare the energy input from radio galaxies
in groups and clusters, we include in our analysis a sample of well
known radio galaxies in massive clusters, extracted from the sample of
(\citealt{birzan04}; see Section \ref{birzan}).  We use their tabulated value of $p\mathrm{d}V$ to compute the
mechanical energy input over the average time the galaxy has spent as a radio
galaxy. Birzan et al.\ provide a value for the energy input for both
filled and radio ghost cavities. In order to obtain a measure
of the average input, we sum the $pdV$ for all the cavities in a
cluster and multiply it by the number of events (i.e.\ how often the radio jet was turned on). The latter
is given by the ratio between $\tau_{radio}$ and the duration of a
single radio event (assuming that all the active AGN phases have the same
duration).

We choose the oldest cavity's age as an indication of the duration of
the radio event. \citet{birzan04} calculate the age of each cavity
in three ways: 1. the time required for the cavity to rise at the sound velocity;
2. the time required for the bubble to rise buoyantly at the terminal velocity;
3. the time required to refill the displaced volume.
We adopt the average of the three age estimates; this is generally similar to the age computed for a buoyantly rising
bubble. We take the error on the cavity's age to be the difference
between the shortest and longest life--time estimated via the three
different methods.

As $\tau_\mathrm{radio}$ for our sample is derived following \citet{smolcic09}, it depends on the redshift
  and the stellar mass of the radio galaxy.  We computed the stellar
  masses for the central radio galaxy in the massive cluster sample
  using the K--band photometry provided by the 2MASS survey
  \citep{2mass}. This method is robust, since radio galaxies contain mostly type~2 (obscured) AGN, whose emission does not significantly contaminate the optical--NIR part of the galaxy spectrum.
   We assume a M/L$_{K}$ ratio for a stellar population
  with an age of $\sim$10 Gyr (corresponding to the age of the stars
  in a galaxy at z$\sim$0), obtained by \citet{drory04}
  (M/L$_{K}$=1.4 with a Salpeter IMF). The quoted error on the M/L$_{K}$ in \citet{drory04} is 25--30$\%$: a change in stellar mass of this magnitude does not affect significantly the time--scales
  we estimate.
  The stellar masses are then
  converted to a Chabrier IMF by subtracting an offset of 0.2 dex.
  
 \subsection{Binding energy of the intra--cluster medium} 
We compute the binding energy for the Birzan et al.\ clusters in the same way as for the COSMOS groups, using the value of M$_{200}$ and R$_{200}$ provided by the X--ray analysis in the HIFLUGCS survey (Reiprich $\&$ Boehringer 2002). We assume a constant concentration parameter of 5. Errors on $E_\mathrm{binding}$ are propagated numerically via a Monte Carlo method, in the same way as the COSMOS groups (see Section \ref{binding}).
As well as computing the binding energy of clusters individually, we also test the cluster result using the scaling relations adopted for the COSMOS groups, both for computing M$_{200}$ \citep{leauthaud09} and for estimating their mechanical energy output \citep{birzan08}. The change in our calculations does not qualitatively affect our results. The values of $E_\mathrm{binding}$ change by less than a factor 2 on average, while values of $E_\mathrm{mech}$ are perturbed randomly within the error bars.

\section{Results}\label{results}
\subsection{The balance of radio--input and binding energy}
\begin{figure}[!h]
\vspace{-0.1cm}
\begin{center}
\includegraphics[width=\columnwidth, bb=0 100 540 679]{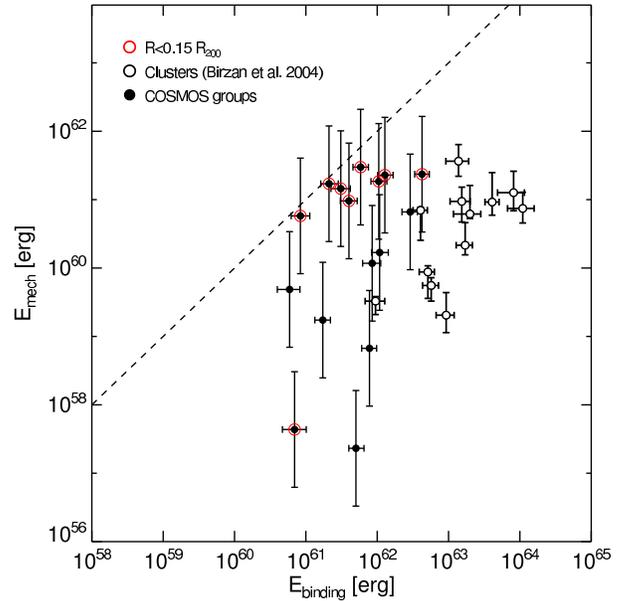}
\caption{Binding energy of the cluster/group gas versus the output
mechanical energy from radio--galaxies. 
Solid points show the 16 X--ray selected groups in
the COSMOS field that host a radio galaxy within their virial radius. 
Large concentric circles indicate groups that host a
radio galaxy within the core (R$\le 0.15\times R_{200}$).  Open points show the
sample of massive local clusters drawn from \citealt{birzan04}. The
dashed line shows equality. { The binding energy in clusters exceeds the total mechanical output by a factor of $\sim$10$^{2}$--10$^{3}$. In all cases except one where a  radio galaxy lies in the centre of a group, the mechanical energy output from the radio galaxy is of the same order as the binding energy for the COSMOS groups analyzed here.}}
\label{fig1}
\end{center}
\end{figure}

Figure~\ref{fig1} shows the binding energy of the gas versus
the energy output from radio galaxies. In the group regime, the
two energies span a comparable range of values (10$^{58}$--10$^{61}$~ergs), 
while for clusters the binding energy exceeds the total
mechanical output of radio--galaxies by a factor approximately of $\sim$10$^2$--10$^{3}$.  { In
particular, for seven groups the two energies are consistent at 1$\sigma$ level, and for all other groups except two the equality holds at 3$\sigma$,
meaning that radio--galaxies potentially provide sufficient energy to
unbind the gas in a large fraction of these groups.}  It is interesting to note that, in all the groups with E$_{mech}\sim$E$_{binding}$, the radio galaxy lies
within 0.15$\times$R$_{200}$ from the center of the group. This suggests that
a radio galaxy in a group is most likely to input sufficient energy into
the ICM to unbind a part of the gas if it lies at the core of the
group. Moreover, radio sources outside the group
core reside in lower density environments, and our calculations of those binding energies
may be overestimates.
The different energy balance in groups and clusters
demonstrates the importance of AGN heating in groups, and shows that
the mechanical removal of gas from groups is energetically possible.
This has important consequences for the understanding
of the baryonic budget in these systems (see \citealt{giodini09}).

\subsection{Can radio galaxies offset radiative cooling in galaxy groups?}
\begin{figure}[!h]
\begin{center}
\includegraphics[width=\columnwidth, bb=0 120 610 649]{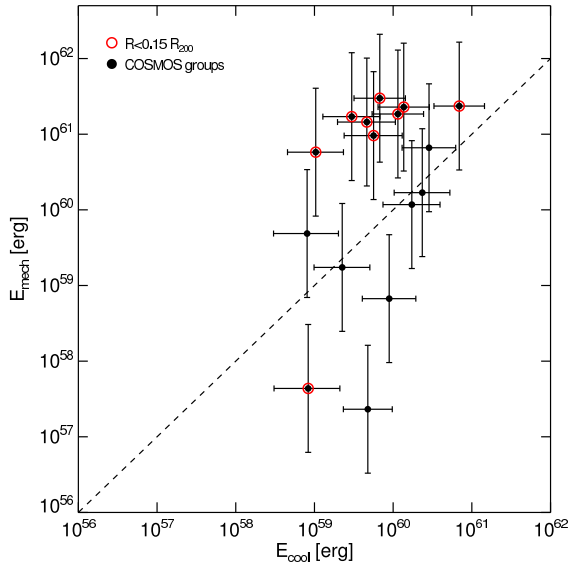}
\caption{The output mechanical energy from radio--galaxies (E$_{mech}$) versus energy radiated inside the cooling radius (E$_{cool}$; i.e. energy
required to offset the cooling in the group center) for 16 X--ray
selected groups from the COSMOS survey . The dashed line shows the equality line. Red
circles mark the radio--galaxies inside $0.15\times R_{200}$. { Uncertainties} on $E_\mathrm{cool}$
are computed allowing an error of a factor 2 on $f_\mathrm{cool}$.}
\label{cooling}
\end{center}
\end{figure}

We now compare the mechanical energy input by radio--galaxies with the
energy required to offset the cooling in the group center
($E_\mathrm{cool}$). As detailed in \citet{fabian94}, \citet{peterson03} and \citet{mcnamara07}, 
the cooling time in cluster/group centers can be lower than the Hubble time, 
implying that large reservoirs of cold gas could accumulate in these regions.
However, evidence that the gas does not cool below approximately one third of the virial temperature \citep{kaastra04}
indicates the presence of a heat source providing enough energy to
offset the cooling. Several studies (e.g. \citealt{peterson03}, \citealt{peterson06}, \citealt{mcnamara07}) suggest AGN feedback
as a viable heating source. To test this hypothesis, we check whether the cooling
energy is { lower than the mechanical energy} of the rising
bubbles.  
We estimate $E_\mathrm{cool}$, assuming that the time during which the gas has been cooling
is equal to the lifetime of the group, which we assume to be 5~Gyr
\citep{voigt04}.  The cooling energy can then be estimated as:
\begin{equation}
E_\mathrm{cool}=L_\mathrm{cool}\times t_{v}=f_\mathrm{cool}\,L_\mathrm{bol}\times t_{v}
\end{equation}
where $t_{v}$ is the lifetime of the group, and
$f_\mathrm{cool}$ is the fraction of bolometric luminosity assumed to be
emitted inside the cooling radius  (where
the cooling time of the gas is lower than the Hubble time).
{ In general this contribution is found to be $\gtrsim$10$\%$ of the total cluster X--ray luminosity \citep{mcnamara07}. Also, the scatter in the L$_{X}$--T scaling relation due to the contribution of cool core clusters can be up to a factor of 2 (\citealt{chen07}; \citealt{pratt09}). 
Given these considerations, we assume that $25\%$ of the total bolometric X--ray luminosity is emitted inside the cooling radius \citep{peres98}.  Since the relative contribution of the cool core to the total X--ray luminosity is higher in groups than in massive clusters, this value is a  good estimate of the average contribution of the cooling core to the total luminosity of a group.}\\
 In Figure~\ref{cooling} we compare
$E_\mathrm{mech}$ and $E_\mathrm{cool}$ in our groups. 
The mechanical energy injected by all but one of the core radio--galaxies
is higher than the radiative losses, and exceeds
$E_\mathrm{cool}$ by an order of magnitude in several cases. We can thus conclude
that radiative losses do not greatly affect the net energy output of
radio--galaxies in the cores of groups. On the other hand, the mechanical output by non-central
radio galaxies is typically of the same order as $E_\mathrm{cool}$.
Moreover, these sources reside mostly outside the cooling radius ($\sim$0.15 R$_{200}$), where the
cooling time is higher than the Hubble time. In this location, the gas does
not lose as much energy through radiative cooling as in the core of the group, so these
galaxies do not provide the required feedback at the right location.
 
\subsection{Impact of systematic effects}
The above calculations rest on several assumptions and should be regarded as rough estimates. 
One critical simplification is the calculation of the
lifetime of a radio--galaxy: the statistical argument used in
\citet{smolcic09} relies on knowledge about the parent
population that hosts the radio--galaxies. In the absence of evidence to the contrary,
we assume that there is no significant
difference between the radio--galaxy elliptical hosts in groups and in low
density environments \citep{feretti07}. We note that even if $\tau_\mathrm{radio}$ were incorrect
by a factor of 4, the mechanical output in clusters would still be significantly lower than the binding energy, but would remain consistent with the binding energy for many of the groups (see Figure \ref{fig1}).\\

Other biases may arise from the scaling relation of Birzan et al., which we use to compute the mechanical energy: { the large scatter in the P$_{1.49\,GHz}$--P$_{cav}$ relationship (0.85 dex) means that care must be taken when using the inferred value as the mean mechanical energy, since most of our calculations rest on the assumption that over the cluster/group lifetime each burst has on average the same power. Indeed, \citet{nipoti05} suggested that the distribution of the outbursts over the cluster/group lifetime is log--normal rather than gaussian; therefore in any system there would be a good chance of observing smaller than average jet powers. Instead, much of the power would be generated by rare, more powerful outburst, such as that observed in MS 0735+7421 by \citet{gitti07}.
These arguments rest on the assumption that the observed scatter in P$_{1.49\,GHz}$--P$_{cav}$ in the observed ensemble of clusters is a good description of the time variability of the AGN power in individual objects.
 In general the ensemble scatter is an upper limit to the scatter in the time variability. If we assume this scatter to represent also for the time variability, we are statistically underestimating the mechanical energy output over the group lifetime by a factor that we compute as follows. The scatter in the Birzan et al. relationship (0.85 dex) corresponds to a probability $\gtrsim$80$\%$ of observing a value smaller than the mean from a single observation (cf. \citealt{nipoti05}). Thus, if we assume that the observed value of  P$_{1.49\,GHz}$ scatters around the median of the distribution, the ratio between the median and the mean for a lognormal distribution (which depends only on the scatter $\sigma$) tells us the scaling factor for the 'true' mean mechanical energy:
\begin{equation}
\frac{mean}{median}=\frac{e^{\sigma^2}}{2}=6.8.
\end{equation}

Therefore the typical observed mechanical power may be underestimated by a factor $\le$7 with respect to the mean. This value, though not negligible, goes in the direction of further increasing the mechanical output, confirming the effect we found.\\}
Furthermore, if the bubble
were over--pressured { when compared to the surrounding ICM \citep{heinz98}, the expanding bubble would carry a shock and the mechanical power may be underestimated, as well as reported by  \citet{birzan04}. }This effect would also boost the mechanical energy to higher values, further strengthening our results.

We have also used preliminary results from VLA 324 MHz data (Smolcic et al.\ in preparation) to double-check our estimates of the mechanical energy output from radio--galaxies. Only 12 of the 16 radio--galaxies are detected in the 324 MHz band and, in all these cases, $E_\mathrm{mech}$ computed using these data (using Eq.15 in \citealt{birzan08})
is consistent within the error bars with the value computed at 1.49 GHz.
As a further check, the total radio luminosity can be computed with higher precision from
break frequencies for 7 of the 16 sources, using the the \citet{myers85} approximation.
The value of $E_\mathrm{mech}$ obtained with this improved method is consistent within the error bars
with that obtained using monochromatic data.

\section{Discussion: the entropy in X--ray groups}\label{discussion}
The injection of energy by radio galaxy activity into the ICM modifies the
thermodynamical state of the gas, raising the entropy ($S$) by a significant
amount compared to that generated by gravitational collapse. { We define the entropy as
\begin{equation}
S\propto \frac{kT}{n_{e}^{\frac{2}{3}}}
\end{equation}
where T is the gas temperature in keV and n$_{e}$ is the gas electron density \citep{voit05}.}
An excess entropy of 50--100 keV cm$^{-2}$ is indeed observed at the group regime,
 causing a deviation from the $S$--T relation \citep{ponman03}. 
 { The excess entropy is measured in the central regions (at 0.1 R$_{200}$).}
In the following we make an order of magnitude calculation of
the excess entropy generated by the energy injected, then compare it
with that observed in groups and predicted from the
theory. We recompute the expected S--T relation taking into account this excess energy, 
and compare it with the observational constraints of \citet{ponman03}.\\

The change in entropy caused by injection of energy under constant
pressure is
\begin{equation}
\Delta S=\frac{2}{5}\frac{\Delta E}{n_{e}}\frac{\gamma_{T}^{5/3}-1}{\gamma_{T} - 1}
\end{equation}
\citep{lloyd00}, where $\Delta E$ is the injected energy per
particle, $\gamma_{T}$ is the ratio between the initial and final
temperature (a value between 1.1 and 2.0, \citealt{lloyd00}), and $n_{e}$ is
the initial electron density ($n_{e}$=10$^{-2}$ { assuming the energy is deposited in the cluster core}, e.g. \citealt{sanderson03}).  We compute
$\Delta E$ from the mechanical energy input of radio galaxies as follows:
\begin{equation}
\Delta E=E_\mathrm{mech}\times \frac{m_{p}\mu}{M_{gas}}
\end{equation}
with $M_\mathrm{gas}=f_\mathrm{gas}\,M_{200}$, where $f_\mathrm{gas}$ is estimated from the
relation between gas fraction and total mass in \citet{pratt09}.  This
calculation does not depend on the details of the energy injection
process.  We obtain values of excess entropy between 10 and 60~keV~cm$^{-2}$. 
This is a rough calculation but predicts values similar to those in \citet{voit05b}. These authors
show that 
an additional energy input episodic on $10^{8}$ yrs timescale is needed
to explain the excess entropy found observationally in the core of clusters \citep{ponman03,donahue05}. 
The additional energy produces an entropy pedestal: \cite{voit05} calculates 10 keV cm$^{-2}$ to be the minimum entropy boost needed to explain observations, and he predicts it to be larger for groups.

The mechanical energy injected by radio galaxies into the 16 COSMOS X-ray selected groups is roughly independent on the group mass (see Figure \ref{fig1}). This is
not unexpected, since the black--hole masses (which are a
zeroth order indicator of the mechanical energy output; \citealt{merloni07}) range only 
between 10$^{8}$--10$^{9}$ M$_{\odot}$ in radio galaxies (see Figure 7 in \citealt{smolcic09}). 
{  At the cluster regime may not be true that the mechanical energy is independent on cluster mass. Indeed \citet{chen07} infer, from the strength of clusters' cooling cores, that a mechanical input higher than anything observed in groups is necessary to balance the cooling of the gas in the strong cool core clusters. However, it has been shown by the same authors that much ($\sim$90\%) of that energy input would be radiated away to balance the cooling, and therefore would not participate to the mechanical removal of the gas.\\}

\begin{figure}[!h]
\begin{center}
\includegraphics[width=\columnwidth, bb=0 130 609 723]{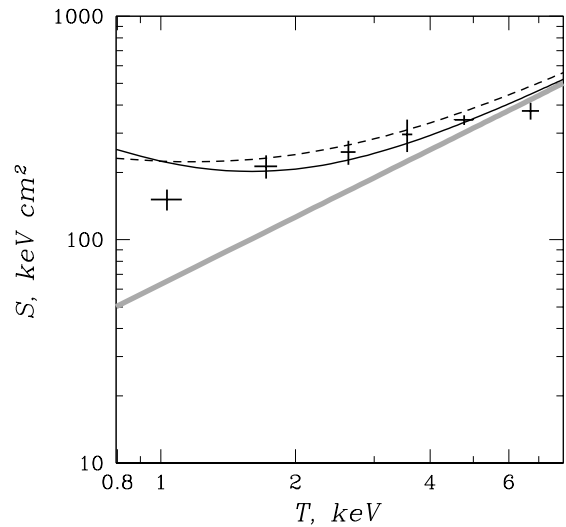}
\caption{Scaling relation between entropy ($S$), measured at 0.1 R$_{200}$, and temperature ($T$). The
  solid black line is the inferred relation accounting for a constant energy
  excess injected by radio galaxies (see text for details). The grey line is the expected
  self--similar relation. The points show { the binned means from the observations} by
  \citet{ponman03}. The dashed line is the same as the solid line but
  considering the self--similar scaling M$_{gas}\propto$T$^{1.5}$}
\label{fun_ex}
\end{center}
\end{figure}

From these considerations, we can
predict how the scaling relation between entropy  and temperature
is affected by the injection of a constant excess energy by
radio galaxies.  As shown in
\citet{finoguenov08}, the energy deposition into the ICM ($\Delta E$)
is proportional to the change in entropy $\frac{\Delta S}{S}$ { for a given typical n$_{e}$. We use  n$_{e}$=10$^{-2}$ as the typical value of the density within 0.1 R$_{200}$, where the majority of the energy is deposited ({  deposition radius};\citealt{sanderson03}).} Using
the scaling of $M_\mathrm{gas}\propto$T$^{2}$ and $E_\mathrm{mech}=\,const$,
then
\begin{equation}\label{dss}
 \frac{\Delta S}{S}\propto \frac{E_\mathrm{mech}}{M_{gas}} \propto \frac{C}{T^{2}}.
\end{equation}
where $C$ is a constant {  and M$_{gas}$ is the mass of the gas within the deposition radius}. We can then infer the functional dependence of $S$ on the virial temperature of the ICM as
\begin{equation}
S=S_{0}+\Delta S=S_{0}\times(1+\frac{\Delta S}{S_{0}})\propto (T_{0}+\frac{C}{T_{0}}),
\end{equation}
where $S_{0}$ and T$_{0}$ are respectively the entropy and the temperature of the gas before the injection of energy from a radio galaxy . 
{ The value of $C$ is computed using Equation~3 in \citet{finoguenov08}  and has a median value of  2.56 if the energy is deposited inside the cooling radius. We assume the cooling radius to be 0.10~$R_{200}$ (e.g. \citealt{ponman03}) }
and show the inferred functional form of $S(T)$ in Figure~\ref{fun_ex}. 
Remarkably, the shape of the resulting scaling relation
(solid line) deviates from the self--similar one (dashed line)
around $\sim$4~keV, in agreement with the observed scaling relation measured at 0.1 R$_{200}$ by \citet{ponman03} (black crosses; { these points are binned means}).  {The deviation of the $\sim$1 keV point indicates that a lower excess entropy is needed to explain very cold groups. This can be achieved requiring that the mechanical energy is deposited at a larger radius in these groups. {  Indeed if the deposition radius increases also M$_{gas}$ within this radius increases. Therefore using Equation \ref{dss} we would obtain a lower values of $\frac{\Delta S}{S_{0}}$ (and thus entropy) for these groups, matching eventually the observational point of \citet{ponman03} at $\sim$1 keV;} if this is the case, it would  confirm that the effect of feedback is more global in groups than in clusters \citep[cf.]{deyoung10}.}
Therefore, the injection of an excess
energy that is independent of groups' mass, thus temperature, (as we
observe from radio galaxies in the COSMOS groups) correctly predicts the
deviation of the observed $S$--$T$ relation from the purely gravitational relation at the group scale.

\section{Conclusions}
 In this Paper we have quantified the importance of the mechanical
 energy input by radio galaxies inside galaxy groups. In particular we
 report a striking difference between clusters and groups of galaxies:
 while the binding energy of the ICM in clusters exceeds the mechanical
 output by radio AGN, the two quantities are of the same order
 of magnitude in groups { that host a radio galaxy within 0.15 R$_{200}$}. This suggests that, while clusters can be mostly
 considered to be closed systems, the mechanical removal of
 gas is energetically possible from groups.
 This has implications that help explain recent findings on the baryonic fraction in
 groups of galaxies. \citet{giodini09} reported a $\sim$30$\%$
 lack of gas in groups compared with the cosmological baryon mass
 fraction evaluated from the 5 years \textit{Wilkinson Microwave Anisotropy Probe} \citep{dunkley08}. 
 It has been suggested that this gas has been removed by AGN feedback.

 This is consistent with cosmological models in which feedback from radio galaxies is invoked to successfully explain galaxy group/cluster properties.
 Based on a well selected sample of galaxy groups and clusters that host radio galaxies,  we have observationally shown for the first time that this scenario is energetically feasible.
 We have further shown that a constant injection of excess energy by radio galaxy naturally reproduces the self-similar breaking observed in the scaling relation between the entropy and temperature of groups.

\acknowledgements
SG acknowledges support by the DFG Cluster of Excellence `Origin and
Structure of the Universe' (\url{www.universe-cluster.de}). SG thanks A. Merloni for helpful comments. 
VS acknowledges support from the Owens Valley Radio Observatory,
which is supported by the National Science Foundation through grant AST-0838260. 
VS and AO thank Unity through Knowledge Fund (www.ukf.hr) for collaboration support through the "Homeland Visit" grant.
AO thanks California Institute of Technology for generous support through NASA grants 1292462 and 1344606.
We acknowledge the contributions of the entire COSMOS collaboration; more informations on the COSMOS survey are available at \url{http://www.astr.caltech.edu/$\sim$cosmos}.

\clearpage
\appendix
\section{Radio Images}
\begin{figure}[!h]
\caption{Figure \protect\ref{fig4} presents the contours maps of the radio 20 cm emission (magenta lines) superimposed to the SUBARU  $zp$ band image for each of the groups listed in Table \ref{tab1}. 
Images are 3$\times$3 arcmin and centered on the group center, except XID246 which is 4$\times$4 arcmin wide and offset from the center group because of its location on the edge of the SUBARU field coverage.
The white line shows the contours of X--ray flux significance. The contours correspond to $\left[3,6,9,12,15,18,21,24\right]\,\sigma$ X--ray flux significance.
}
\begin{center}
$\begin{array}{c@{\hspace{1in}}c}
\multicolumn{1}{l}{\mbox{}} &
	\multicolumn{1}{l}{\mbox{}} \\ 
\includegraphics[width=3.3in]{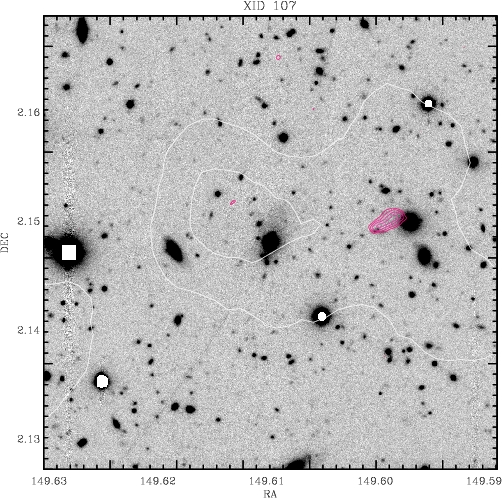} &
 \hspace{-2cm}
\includegraphics[width=3.3in]{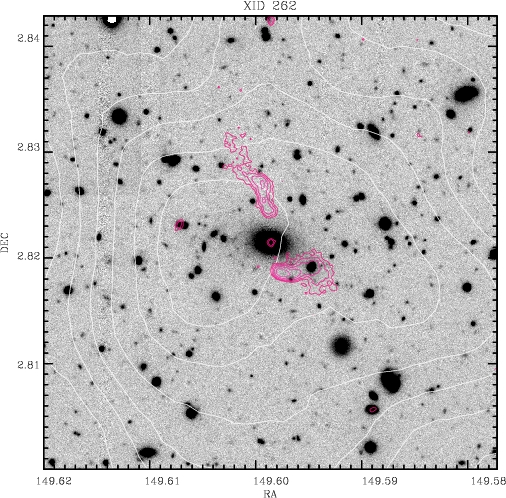} \\ [0.4cm]
\multicolumn{1}{l}{\mbox{}} &
	\multicolumn{1}{l}{\mbox{}} \\ 
\includegraphics[width=3.3in]{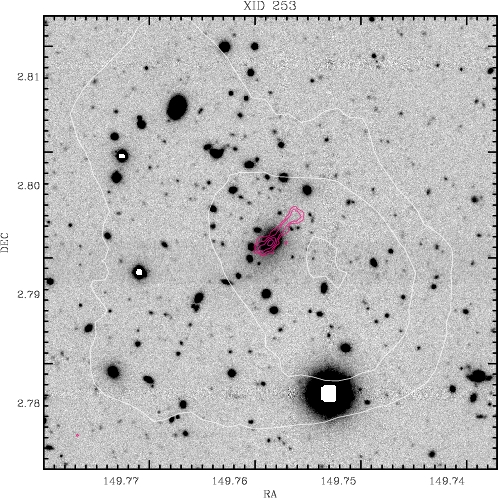} &
\hspace{-2cm}
\includegraphics[width=3.3in]{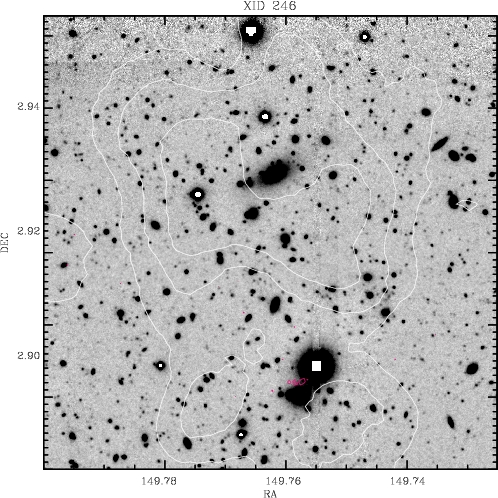} \\ [0.4cm]
\end{array}$
\end{center}
\label{fig4}
\end{figure}

\newpage
\begin{figure*}[h]
\begin{center}
$\begin{array}{c@{\hspace{1in}}c}
\multicolumn{1}{l}{\mbox{}} &
	\multicolumn{1}{l}{\mbox{}} \\ 
\includegraphics[width=3.3in]{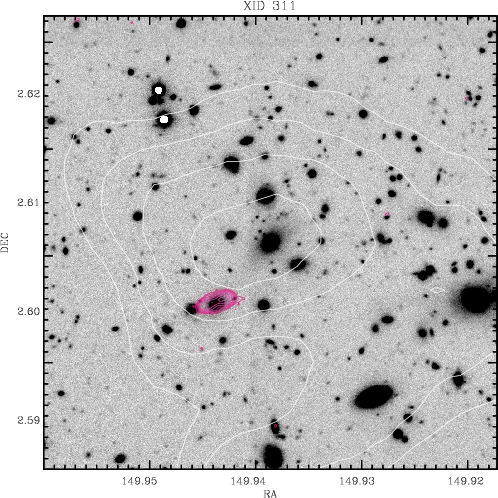} &
\hspace{-2cm}
\includegraphics[width=3.3in]{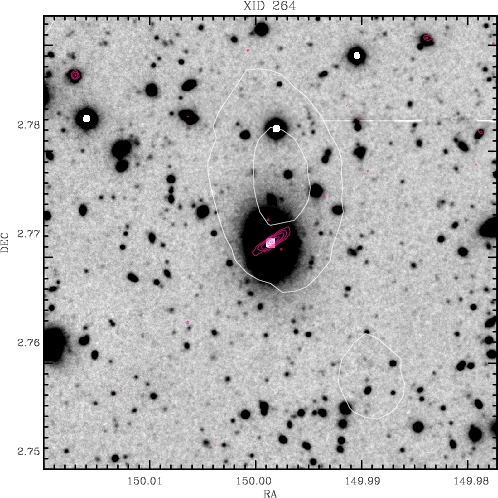} \\ [0.4cm]

\multicolumn{1}{l}{\mbox{}} &
	\multicolumn{1}{l}{\mbox{}} \\ 
\includegraphics[width=3.3in]{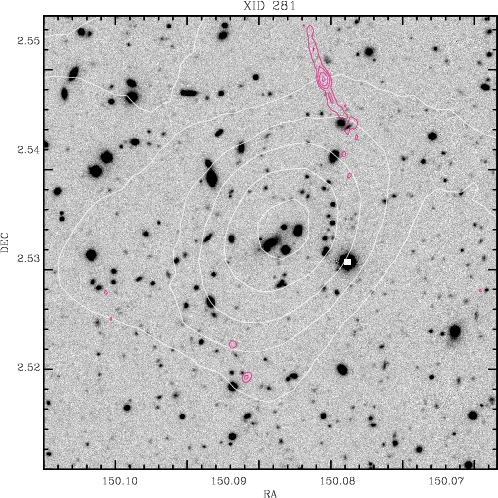} &
\hspace{-2cm}
\includegraphics[width=3.3in]{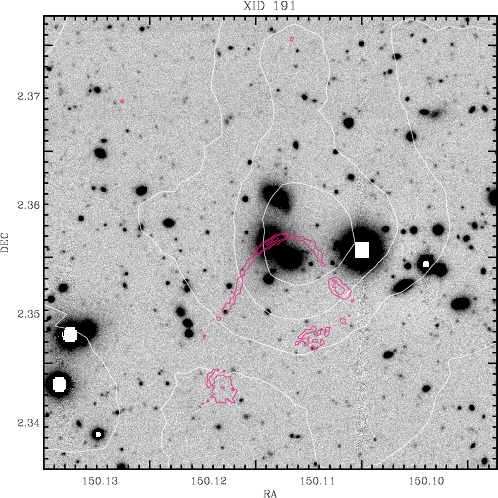} \\ [0.4cm]

\end{array}$
\end{center}
\end{figure*}

\newpage
\begin{figure*}[h]
\begin{center}
$\begin{array}{c@{\hspace{1in}}c}
\multicolumn{1}{l}{\mbox{}} &
	\multicolumn{1}{l}{\mbox{}} \\ 
\includegraphics[width=3.3in]{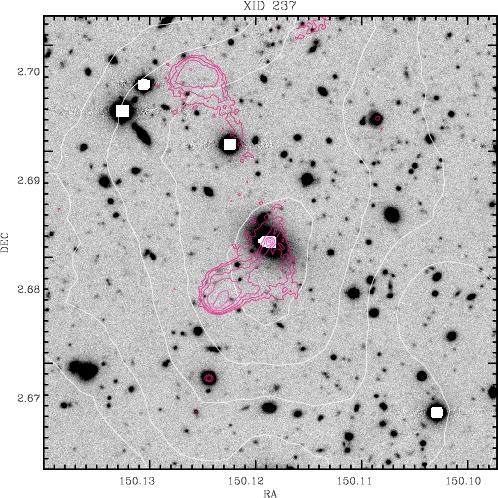} &
\hspace{-2cm}
\includegraphics[width=3.3in]{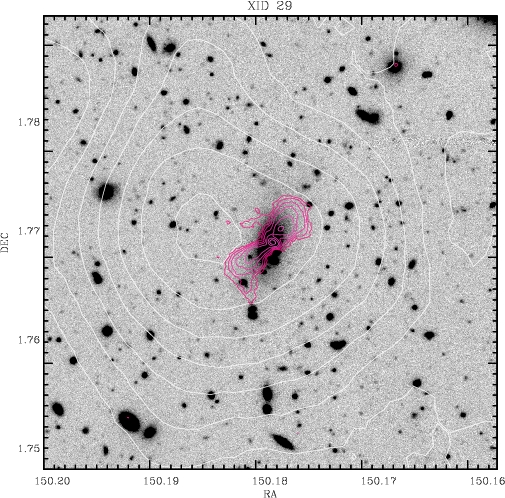} \\ [0.4cm]

\multicolumn{1}{l}{\mbox{}} &
	\multicolumn{1}{l}{\mbox{}} \\
\includegraphics[width=3.3in]{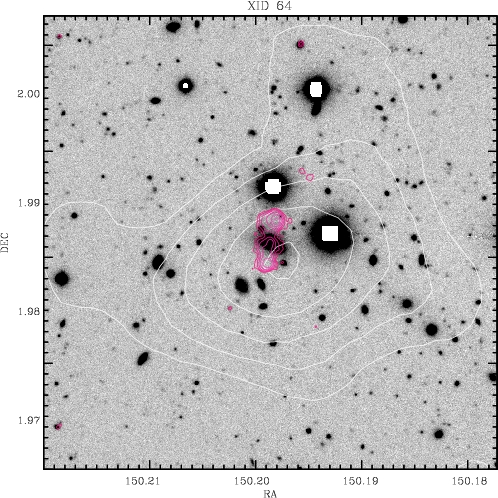} &
\hspace{-2cm}
\includegraphics[width=3.3in]{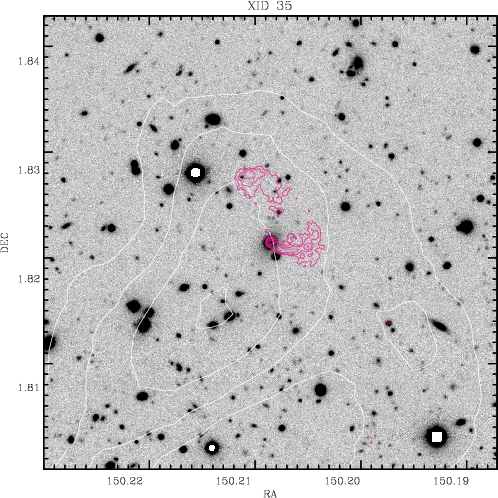} \\ [0.4cm]

\end{array}$
\end{center}
\end{figure*}

\newpage
\begin{figure*}[h]
\begin{center}
$\begin{array}{c@{\hspace{1in}}c}
\multicolumn{1}{l}{\mbox{}} &
	\multicolumn{1}{l}{\mbox{}} \\ 
\includegraphics[width=3.3in]{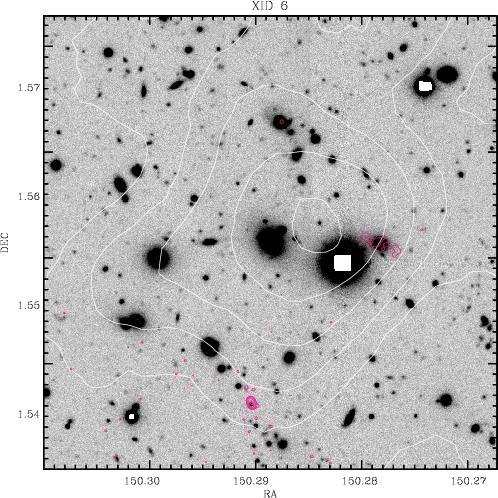} &
\hspace{-2cm}
\includegraphics[width=3.3in]{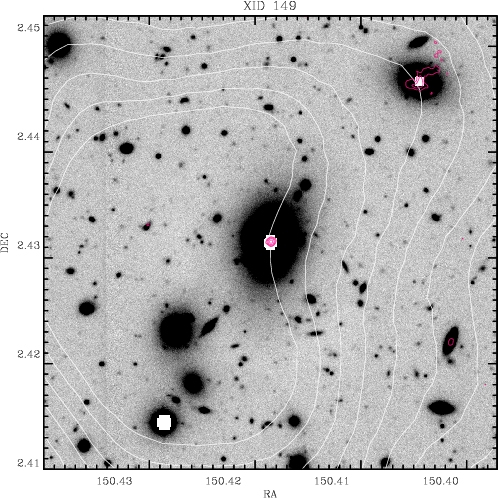} \\ [0.4cm]

\multicolumn{1}{l}{\mbox{}} &
	\multicolumn{1}{l}{\mbox{}} \\ 
\includegraphics[width=3.3in]{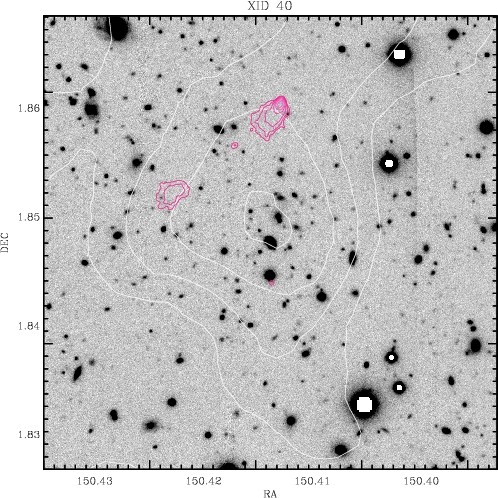} &
\hspace{-2cm}
\includegraphics[width=3.3in]{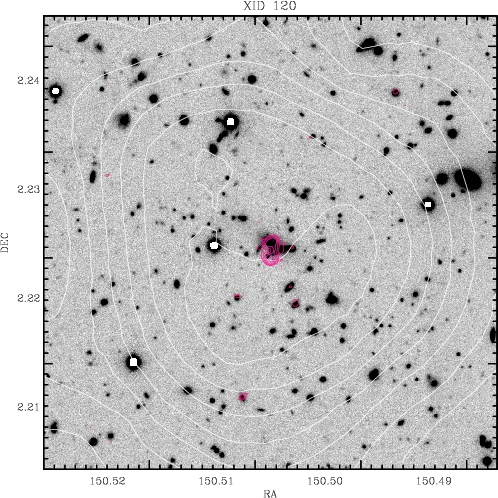} \\ [0.4cm]

\end{array}$
\end{center}
\end{figure*}

\clearpage

\end{document}